\documentstyle[aps,epsfig,prl]{revtex}

\def\Journal#1#2#3#4{{#1} {\bf #2}, #3 (#4)}

\def\PLB{{ Phys. Lett.}  B}
\def\PRL{Phys. Rev. Lett.}
\def\PRC{{Phys. Rev.} C}
\def\PRD{{Phys. Rev.} D}

\def\be{\begin{equation}}
\def\ee{\end{equation}}
\def\bea{\begin{eqnarray}}
\def\eea{\end{eqnarray}}

\newcommand{\kl}{\mbox{$K_L$}}

\newcommand{\epl}{\mbox{$e^+$}}
\newcommand{\emi}{\mbox{$e^-$}}

\newcommand{\pim}{\mbox{$\pi^-$}}

\newcommand{\piz}{\mbox{$\pi^0$}}
\newcommand{\pizd}{\mbox{$\pi^0_D$}}

\newcommand{\h}{\mbox{$H^0$}}
\newcommand{\cas}{\mbox{$\Xi^0$}}
\newcommand{\lam}{\mbox{$\Lambda$}}
\newcommand{\mh}{\mbox{$M_H$}}
\newcommand{\ml}{\mbox{$M_\Lambda$}}

\newcommand{\mc}{\mbox{$M_\Xi$}}

\newcommand{\gev}{\mbox{$\rm{GeV/c^2}$}}
\newcommand{\mev}{\mbox{$\rm{MeV/c^2}$}}

\newcommand{\beq}{\begin{eqnarray}}
\newcommand{\eeq}{\end{eqnarray}}

\begin{document}
\draft
% preprint number commands does not seem to work
%\preprint{DRAFT - V1.0}
%\pagestyle{plain}

%\vspace*{4cm}
\title{
Search for the Weak Decay of a Lightly Bound \h\ Dibaryon
\\
%DRAFT - V2.4  (\today)
}

\author{\parindent=0.in
The KTeV Collaboration\\
A.~Alavi-Harati$^{12}$,
T.~Alexopoulos$^{12}$,
M.~Arenton$^{11}$,
K.~Arisaka$^2$,
S.~Averitte$^{10}$,
A.R.~Barker$^5$,
L.~Bellantoni$^7$,
A.~Bellavance$^9$,
J.~Belz$^{10}$,
R.~Ben-David$^{7,\dagger}$,
D.R.~Bergman$^{10}$,
E.~Blucher$^4$, 
G.J.~Bock$^7$,
C.~Bown$^4$, 
S.~Bright$^4$,
E.~Cheu$^1$,
S.~Childress$^7$,
R.~Coleman$^7$,
M.D.~Corcoran$^9$,
G.~Corti$^{11}$, 
B.~Cox$^{11}$,
M.B.~Crisler$^7$,
A.R.~Erwin$^{12}$,
R.~Ford$^7$,
A.~Glazov$^4$,
A.~Golossanov$^{11}$,
G.~Graham$^4$, 
J.~Graham$^4$,
K.~Hagan$^{11}$,
E.~Halkiadakis$^{10}$,
K.~Hanagaki$^8$,  
S.~Hidaka$^8$,
Y.B.~Hsiung$^7$,
V.~Jejer$^{11}$,
D.A.~Jensen$^7$,
R.~Kessler$^4$,
H.G.E.~Kobrak$^{3}$,
J.~LaDue$^5$,
A.~Lath$^{10}$,
A.~Ledovskoy$^{11}$,
P.L.~McBride$^7$,
A.P.~McManus$^{11}$,
P.~Mikelsons$^5$,
E.~Monnier$^{4,*}$,
T.~Nakaya$^7$,
K.S.~Nelson$^{11}$,
H.~Nguyen$^7$,
V.~O'Dell$^7$, 
M.~Pang$^7$, 
R.~Pordes$^7$,
V.~Prasad$^4$, 
C.~Qiao$^4$,
B.~Quinn$^4$,
E.J.~Ramberg$^7$, 
R.E.~Ray$^7$,
A.~Roodman$^4$, 
M.~Sadamoto$^8$, 
S.~Schnetzer$^{10}$,
K.~Senyo$^8$, 
P.~Shanahan$^7$,
P.S.~Shawhan$^4$,
W.~Slater$^2$,
N.~Solomey$^4$,
S.V.~Somalwar$^{10}$, 
R.L.~Stone$^{10}$, 
I.~Suzuki$^8$,
E.C.~Swallow$^{4,6}$,
S.A.~Taegar$^1$,
R.J.~Tesarek$^{10}$, 
G.B.~Thomson$^{10}$,
P.A.~Toale$^5$,
A.~Tripathi$^2$,
R.~Tschirhart$^7$, 
Y.W.~Wah$^4$,
J.~Wang$^1$,
H.B.~White$^7$, 
J.~Whitmore$^7$,
B.~Winstein$^4$, 
R.~Winston$^4$, 
T.~Yamanaka$^8$,
E.D.~Zimmerman$^4$\\
\vspace*{0.1in}
\footnotesize
$^1$ University of Arizona, Tucson, Arizona 85721 \\
$^2$ University of California at Los Angeles, Los Angeles, California 90095 \\
$^{3}$ University of California at San Diego, La Jolla, California 92093 \\
$^4$ The Enrico Fermi Institute, The University of Chicago, 
Chicago, Illinois 60637 \\
$^5$ University of Colorado, Boulder, Colorado 80309 \\
$^6$ Elmhurst College, Elmhurst, Illinois 60126 \\
$^7$ Fermi National Accelerator Laboratory, Batavia, Illinois 60510 \\
$^8$ Osaka University, Toyonaka, Osaka 560 Japan \\
$^9$ Rice University, Houston, Texas 77005 \\
$^{10}$ Rutgers University, Piscataway, New Jersey 08855 \\
$^{11}$ The Department of Physics and Institute of Nuclear and 
Particle Physics, University of Virginia, 
Charlottesville, Virginia 22901 \\
$^{12}$ University of Wisconsin, Madison, Wisconsin 53706 \\
$^{*}$ On leave from C.P.P. Marseille/C.N.R.S., France \\
$^\dagger$ To whom correspondence should be addressed. Electronic
address: rbd@fnal.gov\\
\normalsize
}

\maketitle

\begin{abstract}
We present results of a search for a neutral, six-quark, dibaryon
state called the \h, 
 a state predicted to exist in several theoretical
models.   Observation of such a state would signal the discovery of 
a new form of hadronic matter. 
Analyzing data collected by experiment  E799-II, using the KTeV detector 
at
Fermilab, we searched for the decay  $H^0 \rightarrow \Lambda p
\pim$ and found no candidate events. 
We exclude the  region 
 of lightly bound mass states just below the $\Lambda\Lambda$
mass threshold, $2.194~\gev < \mh < 2.231$~\gev,
with lifetimes from $\sim$$ 5 \times 10^{-10}$~sec to 
$\sim$$1 \times 10^{-3}$~sec.
\end{abstract}


%12. Specific Theories and Interaction Models; Particle Systematics
%13. Specific reactions and phenomenology
%14. Properties of specific particles
%21. Nuclear Structure 

%14.20.Pt Dibaryons
%13.85.Rm Limits on production of particles
%13.25.Ev Hyperon-nucleon interactions
%12.39.Ba Bag model
%21.80.+a Hypernuclei

\twocolumn  

%To date, the only hadrons that have been observed are  mesons
%($q\bar{q}$) and baryons ($qqq$).  
In 1977, Jaffe~\cite{jaffe} proposed the existence of a metastable
dibaryon, the \h($hexa$-quark), a 
bound six-quark state ($B=2$, $S=-2$), described as $\h=|uuddss\rangle$.
If it exists, this  hadron would  be a 
 new form of matter.
The observation
of a bound dibaryon would enhance
the understanding of strong interactions and would aid
in the search for additional exotic-multiquark 
states~\cite{lipkin1}.

The two-flavor six-quark
state is unbound~\cite{lipkin2},
a result of the  Pauli exclusion principle. The Pauli exclusion
principle
  can be circumvented through the addition
of strangeness as an
extra degree of freedom.  
Jaffe estimated that the color-hyperfine interaction between the
six quarks of a $|uuddss\rangle$ state 
would be strong enough to cause the \h\ to be a bound state.
Different theoretical models have produced a multitude
of predictions for \mh, covering a broad mass range from deeply bound
states to unbound states~\cite{quinn}.  Most of the predictions, however, 
 are clustered in the range of
2.1~\gev\ up to a few \mev\ above the $M_{\Lambda \Lambda}$ threshold of
2.231~\gev. 
If \mh\ is between the $M_{\Lambda n}$ (2.055~\gev)
 and $M_{\Lambda \Lambda}$ thresholds, it
is expected to be a metastable state and undergo a $\Delta S = 1$ weak decay.
Its lifetime is estimated to be less than $\sim$$2 \times
10^{-7}$~sec~\cite{donoghue}, while  baryonic  $\Delta S = 1$ weak
decays suggest a lower limit on the lifetime of $\sim$$1 \times 10^{-10}$~sec.

Using a
 variety of techniques,  experimentalists 
have been trying for
years to detect the \h, without conclusive results~\cite{ashery}.  In recent
years, production models based on empirical data
with few assumptions built into them 
 have allowed experimentalists to
gauge the sensitivity of their results.  In particular, the combined
results from three recent experiments~\cite{ashery,ahn,belz}
 rule out the mass range for
 $\Delta S = 1$ transitions below
2.21~\gev. 
The analysis presented here covers the mass
range of lightly bound \h's between the $M_{\Lambda p \pi^-}$  and
$M_{\Lambda\Lambda}$ thresholds, 2.194~\gev and 2.231~\gev,
respectively.
In addition, this search is
 sensitive to a large range of lifetimes, from  
$\sim$$5\times 10^{-10}~{\rm sec}$ to $\sim$$1 \times 10^{-3}$~sec, completely 
covering the
range of lifetimes
proposed in
reference~\cite{donoghue} yet to be probed. 

It is expected that an \h\ can be produced in $pN$ collisions through
hyperon production; where two strange quarks are produced, followed by
the coalescence of a hyperon and a baryon to form a bound six-quark
state.
Currently, the only model for \h\ production at Tevatron beam energies
is one proposed by Rotondo~\cite{rotondo}.
His model is based on production of the doubly strange \cas, 
 followed by the coalescence of
the \cas\ with a $n$, predicting a total cross-section of 1.2~$\mu$b.
Our search for the \h\ is the first  search to normalize to the 
doubly strange \cas, removing  the strangeness production portion from
the \h\ production process, 
 making this analysis a sensitive probe of hypernuclear coalescence.
The \h\ production process at the Tevatron, through $pN$
collisions, complements other current experimental  efforts  that search for
\h's produced in heavy ion collisions.

%heavy ion beams
%is 
% makes 
%this production process  complementary to current searches being carried out using 
%lower energy 
%heavy ion beams.  Clean beam conditions allow us to fully
%reconstruct the final state.

The KTeV beam line and detector at Fermilab were designed
for high precision studies of direct CP violation in the neutral kaon system
(E832) and in  rare \kl\ decays (E799-II).  To reduce backgrounds from long
lived neutral states,  the apparatus was situated far
from the production target.
A clean neutral beam, powerful particle identification, and very good
resolution for both charged particles and photons made it a good
facility to search for and fully reconstruct 
both the signal mode,  $\h \rightarrow
\Lambda p \pim$,  and the normalization mode,
$\cas \rightarrow \Lambda \pizd$, where \pizd\ refers to the Dalitz
decay of the \piz\ to
$\epl\emi\gamma$. 
For both modes, the $\Lambda$'s decay downstream of the parent
particle's vertex to a $p\pim$.
%The topologies for both modes are similar. 
%The parent particle's decay vertex is defined by the charged track vertex of 
%the $p \pim$, for the \h, or the \epl\emi, for the \cas.  Downstream 
%of the  parent particle's decay vertex is the $\Lambda$'s decay
%vertex, which decays to $p\pim$.
The data presented here were collected during two months of E799-II
data-taking in 1997.

\begin{figure*}[htb]
\vspace*{-4em}
\centerline{\epsfig{height=3.9in, angle=270., file=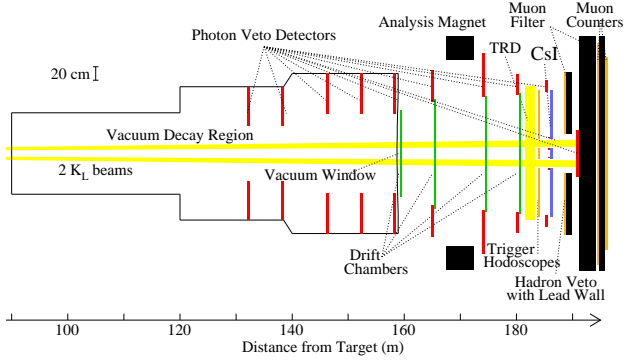}}
\vspace*{-0.2in}
\caption{Plan view of the KTeV spectrometer.}
\label{fig:detector}
\end{figure*}

The KTeV detector and the trigger configuration used to select events
with four charged particles have been described elsewhere~\cite{4track}.
This article highlights aspects of the
detector directly relevant to this analysis.
A neutral beam, composed primarily of kaons and neutrons was produced
by focusing an 800~GeV/$c$ proton beam 
at a vertical angle  of 4.8 mrad  on a 1.1 interaction length
(30~cm) BeO target. Photons produced in the target were converted in a 
7.6~cm  lead absorber located downstream of the target. 
Charged  particles were removed further downstream 
with magnetic sweeping.
Collimators, followed by sweeping magnets, 
 defined two 0.25~$\mu$sr neutral beams
that entered the KTeV apparatus (Fig.~\ref{fig:detector})
94~m downstream from the target. The 65~m   vacuum
($\sim$$10^{-6}$~Torr) decay region  extended  to the   first   drift
chamber.
 
The momenta of the charged particles were measured with a
 charged particle spectrometer, consisting of four planar drift chambers,
two upstream and two downstream of a dipole analyzing magnet.
The energies of  the particles were measured with a high resolution 
CsI  electromagnetic calorimeter.
To distinguish electrons from hadrons, the energy (E) measured by
the calorimeter was compared to the momentum (p) measured by the charged
spectrometer. Electrons were identified by $ 0.9 < {\rm E/p} < 1.1 $,
while pions and protons were identified by ${\rm E/p} < 0.9$.

Offline, events were required to have four reconstructed charged
  particles.
We searched for long lived \h's which were produced at the target and 
 decayed in the vacuum decay region.  
A characteristic feature of the topology 
of both the signal and normalization modes
 is that the parent 
particle's true decay vertex is defined  by a charged track vertex;
the $p \pim$ vertex for the \h\ and the \epl\emi\ vertex
 for the \cas. The subsequent
$\Lambda$'s decay
downstream 
of the  parent particle's  vertex to  $p\pim$.
Events were required to have at least 
four reconstructed tracks,  two
tracks  associated with positive particles and two with negative
particles.
In the case of the \h, having identified the $p$ and the \pim\ by
their $E/p$ and their charge, there remains a two-fold ambiguity in
combining the $p$'s with the \pim's to form a vertex.  To resolve that
ambiguity, each pair of
 positive and negative
tracks was combined to form a vertex at the location
of 
closest approach  between the two tracks.  
The  distance of closest approach (DOCA) and  
the resultant momentum vector of the combined charged tracks for 
 both
the upstream $p\pim$ vertex and the downstream \lam\ vertex were
calculated.   The \h\
vertex was determined by calculating the DOCA of
 the downstream \lam\ and the upstream $p\pim$.
% constraining the \h\ vertex to be on the vector defined by the 
%upstream vertex.
%The \h\ vertex was constrained to lie along the momentum vector defined 
%by the upstream vertex because the position resolution along the
%direction of the vector is $\sim$50~cm, while the resolution perpendicular  
%to the direction of the vector is $\sim$0.1~cm.
The DOCA's of the upstream, downstream and \h\ vertices were summed
in quadrature, and the permutation that gave the minimum quadrature
sum was selected.

Downstream \lam's were identified by requiring the 
ratio of the lab momenta of the $p$ to \pim\ to be greater
than 3, which accepted 99.8\% of the simulated signal events.
Because the \lam\ decays to two particles, the transverse momentum
($P_T$) distribution
of the decay products relative to the direction of the \lam\
exhibits a Jacobian peak  at a maximum of
0.1~GeV/c.  To enhance the selection of \lam\ decays relative to
background
three body \kl\ decays, where the $P_T$ distribution is
peaked at 0, we required the  $P_T$ of the $p$ and the \pim\ to be between
0.07~GeV/c and 0.11~GeV/c, accepting $\sim$60\% of the  simulated
signal events. 
  To  select \lam's further, we required the
mass of the reconstructed $p\pim$ to fall within $\pm5$~\mev\ of \ml,
where the \ml\ resolution is $\sim$1~\mev.
The charged portion of the upstream  
 vertex is made up of a $p$ and \pim\ and has
kinematics similar to a \lam\ decay.
Thus,
the same  constraint used for the \lam\ was applied,
 requiring the ratio of the lab 
momenta of the $p$ to \pim\ to be greater than 3.

Interactions in the  collimator,  sweeping magnets, 
and vacuum window produced background 
events with multiple vertices.  Events where at least 
one decaying  particle was short-lived were
removed by requiring the reconstructed \h\ and \lam\ vertices to be
between 100~m and 155~m from the target,
 $\sim$5~m away from those apparatus elements.

%Additional background events were removed by requiring
% the transverse momentum of 
%the reconstructed \h\
% ($P_T(\h)$), measured relative to a vector connecting the \h\ 
%decay vertex and
%the target, to be less than 0.01~GeV/c.  This cut was set by examining
% the sideband region of   \mh\ in the range $2.235~\gev\ < \mh\ <
%2.280~\gev$ (see Fig.~\ref{fig:ptvmh}).
%Based on our \h\ simulation, we expect $\sim$~80\% of signal events to
%pass the cut on  $P_T(\h)$.

\begin{figure}[!htb]
%\vspace*{-3em}
\centerline{\epsfig{width=3.1in, file=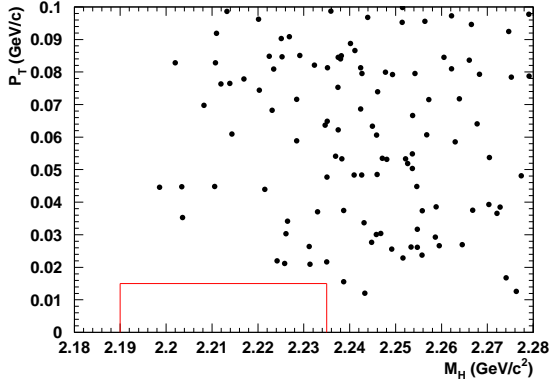}}
\caption{The transverse momentum of the \h\ versus \mh\ for events
passing all the selection criteria except the cuts on $P_T(\h)$ and
\mh. The box shows the signal region. 
The background events outside the signal region are from simultaneous 
decays of two particles in the fiducial volume, the predominant contribution
coming from the decays $\kl \rightarrow \pi^\pm l^\mp \nu$.}
\label{fig:ptvmh}
\end{figure}

The signal region for \h\ candidates was defined by requiring
\mh\ to be between
2.190~\gev\  and 2.235~\gev\ to account for
 resolution effects in  measuring \mh;  the upper and lower 
limits on \mh\  are
  4~\mev\ (more than twice our estimated \mh\ resolution)
 above the $M_{\Lambda\Lambda}$ 
threshold of 2.231~\gev\   and below the $M_{\Lambda p \pi^-}$ threshold 
of 2.194~\gev, respectively.  In addition,  the transverse momentum of 
the reconstructed \h\
 ($P_T(\h)$), measured relative to a vector connecting the \h\ 
decay vertex and
the target was required to be less than 0.015~GeV/c (see Fig.~\ref{fig:ptvmh}).
The cut on $P_T(\h)$ accepted  90\% of the remaining simulated signal events.
None of the events passed all the selection criteria.

\begin{figure}[!htb]
%\vspace*{-3em}
\centerline{\epsfig{width=3.1in, file=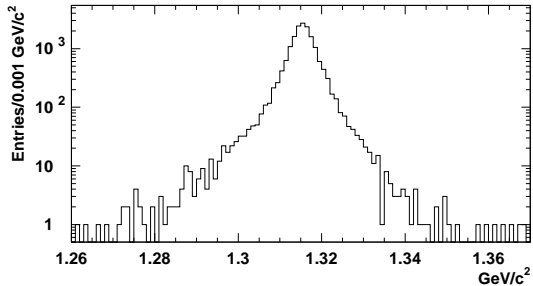}}
\caption{The invariant mass of the $\Lambda \pizd$ sample used to normalize
the signal mode (\mc\ = 1.315~\gev).  There are a total of 17~160 events in the mass peak.}
\label{fig:cas}
\end{figure}

To quantify  the measurement sensitivity,   we
normalized 
to \cas\  production, using data taken with the same trigger
configuration as that 
for the \h\ analysis, reconstructing $\cas \rightarrow \Lambda \pizd$
decays.
Except for the additional photon coming from the \pizd, 
the normalization mode's decay topology   is
similar to that of the \h's. 
Applying a series of cuts similar to those used for the \h\ analysis
 yields 17~160 \cas\ events, with negligible background.
The cleanliness of the normalization mode's signal is
demonstrated in Fig.~\ref{fig:cas} which 
shows the \cas\ invariant mass peak.  
The accepted 
\cas's have mean momentum of $270$~GeV/c.
Distributions of variables from simulated decays,  
such as  the \cas's momentum and the location of 
the \cas's decay vertex are consistant with the same for data.

The \cas\ and \h\ are expected to have different absorption lengths in
the BeO target and the Pb absorber, leading to a difference in the
transmission probability ($T$) for the two particles.
 We estimate the \cas-nucleon ($\sigma_{\Xi N}$) and \h-nucleon
($\sigma_{H N}$) cross-sections and thus the $T$'s
  based on the assumption of isospin invariance.
In addition, we utilize  measured $np$, $\Lambda p$ and
deuteron-proton ($dp$)  
cross-sections~\cite{pdg}, 40~mb, 35~mb and 75~mb, respectively, 
to account for the effect of replacing down
quarks with strange quarks;  we assume that the scale
factor $S = \sigma_{\Lambda p}/\sigma_{np}$ can be used to correct for
the substitution of a single strange quark for a down quark and $S^2$
for a double substitution.  We then estimate $\sigma_{\Xi N}$
 to be $\sigma_{\Lambda p}  S = (31 \pm 4)$~mb and
$\sigma_{H N}$ to be $\sigma_{dp} S^2 = (57 \pm
18)$~mb, where the assigned errors are taken to be equal to the
magnitude of the correction itself.  
The measured absorption lengths for nucleons in BeO and Pb are scaled by 
the  factors $\sigma_{np}/\sigma_{HN}$ and
$\sigma_{np}/\sigma_{\Xi N}$.  The estimated $T$'s in the
target  are $T^{Be0}_\Xi = 0.623 \pm 0.037$ and $T^{Be0}_H = 0.44
\pm 0.12$.  In the lead absorber, the $T$'s
are estimated to be $T^{Pb}_\Xi =
0.562 \pm 0.043$ and $T^{Pb}_H = 0.35 \pm 0.14$.

As  no signal events passed all the selection criteria, the final
result is presented as a 90\% C.L. upper limit on 
the inclusive \h\ production cross-section over the solid angle 
defined by the collimators, expressed in terms of the inclusive \cas\
production cross-section
\beq
\frac{d\sigma_H}{d\Omega} & < &
\frac{\xi}{N_{\Xi}}
\frac{T^{Be0}_\Xi \; T^{Pb}_\Xi}{T^{Be0}_H \; T^{Pb}_H}
 \frac{A_\Xi}{A_H} \frac{B(\cas
\rightarrow \lam \pizd)}{B(\h \rightarrow \lam p
\pim)}\frac{d\sigma_\Xi}{d\Omega},
%\label{eq:summary}
\eeq
where $\xi$ is the factor which multiplies the single event
sensitivity (SES) to give the 90\% C.L. upper limit,
$N_{\Xi}$ is the number of reconstructed $\cas \rightarrow  \lam
\pizd$
decays,
the various $T$ factors are the transmission probabilities described
previously, 
$A_\Xi$ and $A_H$ are the  acceptances for 
\cas\ and \h\ decays, respectively, and $B(\cas \rightarrow \lam
\pizd)$
and $B(\h \rightarrow \lam p \pim)$ are the respective branching
ratios.
Our estimate of the SES
 suffers from a large relative  uncertainty of $\sim$$50\%$,
predominantly due to the  uncertainty in determining the
transmission factors.
The uncertainty in the SES gives rise
to a factor of $\xi = 3.06$ in the determination of the 90\% C.L. upper
limit~\cite{cousins}.

The acceptances were determined  from a detailed detector 
simulation.
Because the trigger was the same for both the signal and normalization 
modes and because both the signal and normalization modes consist of
four-track events with largely similar topologies,
trigger and acceptance
inefficiencies mostly cancel. 
The \cas\ flux was measured using two separate  triggers, each composed of
different trigger elements.  The discrepancy between the two flux
measurements was converted into a systematic  uncertainty in
determining $A_\Xi$.  Other systematic uncertainties were negligible 
relative to this uncertainty.
$A_\Xi$ was determined to be $(6.93 \pm
0.94)\times 10^{-6}$.

\begin{figure}[!htb]
\centerline{\epsfig{width=3.1in, file=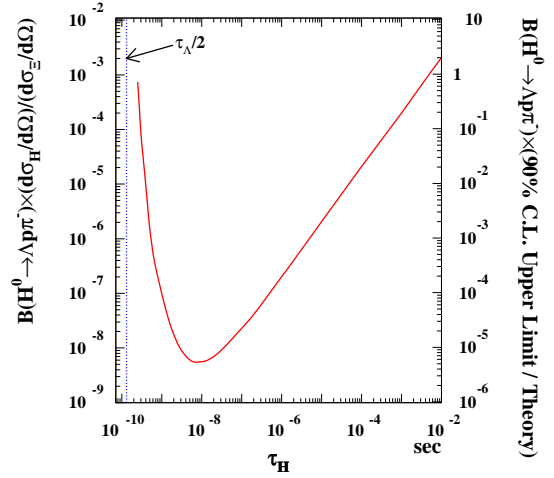}}
\caption{
The 90\% C.L. upper 
limit on the product of the \h\ branching ratio and the production
cross-section as a function of the \h\ lifetime. \mh\ is
assumed to be 
2.21~\gev.
 The
sensitivity scale on the right ordinate axis assumes the production
model of reference~\protect\cite{rotondo}.}
\label{fig:sigvstau}
\end{figure}

To determine $A_H$, the detector simulation 
included the \h\ production spectrum proposed in
Rotondo's phenomenological model~\cite{rotondo}.  
The dominant experimental uncertainty in  $A_H$ comes from the simulation
of proton showers in the calorimeter, where the relative uncertainty
was determined to be
 5.3\%.  
For example, taking \mh\  in the middle of
the mass range we are sensitive to, $\mh =
2.21~\gev$, and  the lifetime corresponding to the
 lifetime given in reference~\cite{donoghue} for this mass,
 $\tau_H = 5.28 \times 10^{-9}$~sec,
$A_H = 5.64\times 10^{-3}$.
As a cross-check of Rotondo's
model,
which incorporates a \cas\ production spectrum,
we applied our measured \cas\ production spectrum in the detector simulation,
replacing  \mc\ and $\tau_\Xi$ with
\mh\ and $\tau_H$, respectively.  This lowered $A_H$ by $\sim$15\%.
 The 90\% C.L. upper limit on the product of the \h\ branching ratio
and the  
production 
cross-section, taking into account all the uncertainties, is 
\beq
B(\h
\rightarrow \lam p \pim) \frac{d\sigma_H}{d\Omega} & <  & 5.87 \times 10^{-9}
\frac{d\sigma_\Xi}{d\Omega}. 
\eeq
In Fig.~\ref{fig:sigvstau}, we plot 
the 90\% C.L. upper limit on the ratio 
$(B(\h\rightarrow \lam p
\pim)d\sigma_H/d\Omega)/(d\sigma_\Xi/d\Omega)$, studying the effect on
$A_H$ of  
varying  $\tau_H$ over a large range of values. For short lifetimes,
the \h's decay before reaching the  decay region,
 while for long lived states,  only a few decay while
passing through the detector.  Both effects lower 
our sensitivity to \h\ decays.
Varying \mh\ across the full range of masses to which  we are sensitive
leads to a relative shift of approximately $\pm60\%$ from the
central value of the
curve plotted in Fig.~\ref{fig:sigvstau}.  Included in the figure is
a line at $\tau_\Lambda/2 = 1.316 \times 10^{-10}$~sec,
 the expected lifetime of system made up of two lightly bound 
$\Lambda$'s, which might be a lower bound on $\tau_H$.
 To interpret the sensitivity of this 
 result relative to the theoretical production model,
 we integrate the theoretical predictions for
 both $d\sigma_H/d\Omega$~\cite{rotondo} and
$d\sigma_\Xi/d\Omega$~\cite{pondrom} 
over the solid angle covered by the collimators.
The right ordinate
axis of Fig.~\ref{fig:sigvstau} shows the sensitivity of this 
measurement.
 Thus our
result rules out lightly bound \h's, between  the
$M_{\Lambda p\pi^-}$
and 
$M_{\Lambda\Lambda}$ thresholds of 2.194~\gev\ and
2.231~\gev, respectively,
over a large range of lifetimes, from $\sim$$5 \times 10^{-10}$~sec up
to $\sim$$1 \times 10^{-3}$~sec.

\begin{figure}[!htb]
\centerline{\epsfig{width=3.1in, file=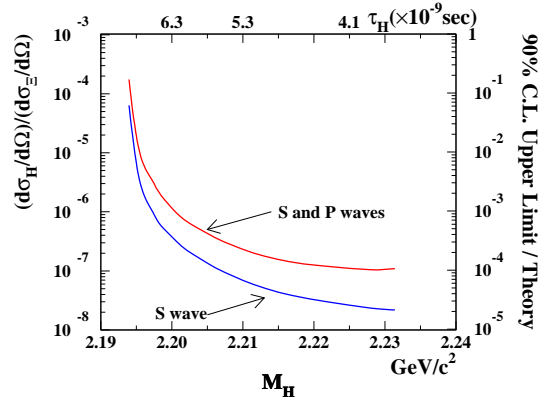}}
\caption{
The 90\% C.L. upper limit on the \h\ production cross-section 
as a function of  \mh.
The top abscissa axis shows the \h\
lifetime associated with
\mh\
assuming the \h's wave function is a pure $S$ wave
state, as predicted in reference~\protect\cite{donoghue}.  If the \h\
contains both $S$ and $P$ wave contributions,
reference~\protect\cite{donoghue}
 estimates the
lifetime could be as low as half the lifetime of the pure $S$ wave state, 
worsening our sensitivity to the \h.  The curve for the $S$ and $P$
wave state is derived using half the lifetime of the pure $S$
wave.
%The dotted lines highlight the upper and lower bounds of the
%mass range we to which we are sensitive, 2.194~\gev\ for the
%$M_{\Lambda p\pi^-}$ threshold and 2.231~\gev\ for the  $M_{\Lambda\Lambda}$ 
%threshold.
}
\label{fig:sigvsmass}
\end{figure}

A model  proposed in reference~\cite{donoghue} associates \mh\ with both
$\tau_H$ and   $B(\h \rightarrow \lam p\pim)$.  For example, for
 $\mh = 2.21~\gev$ they predict 
$\tau_H = 5.28 \times 10^{-9}$~sec and a branching ratio of
$5.4\times 10^{-2}$. 
To test this model, we vary \mh\ between the $M_{\Lambda p \pi^-}$
and $M_{\Lambda\Lambda}$  thresholds,
determining the dependence of the  
production cross-section on the mass, lifetime and branching ratio.
Figure~\ref{fig:sigvsmass} is a plot of 
$(d\sigma_H/d\Omega)/(d\sigma_\Xi/d\Omega)$
 versus \mh.  In this figure, the factor influencing the
sensitivity the most is the \h\ branching ratio which decreases from a
maximum of 14\% at the $M_{\Lambda\Lambda}$ threshold down to
zero at  the
$M_{\Lambda p\pi^-}$ threshold.
The right ordinate
axis of Fig.~\ref{fig:sigvsmass} shows the sensitivity of this 
measurement, based on Rotondo's model.
 Assuming Rotondo's production model, this result clearly
rules out a long lived \h\ state, as 
proposed in  reference~\cite{donoghue}, for \mh\ between the
$M_{\Lambda p\pi^-}$
and 
$M_{\Lambda\Lambda}$ thresholds. 

To conclude, our result rules out  a lightly bound \h\
dibaryon  over a range of mass  below  the $M_{\Lambda\Lambda}$ threshold
not ruled out by previous experiments and for a
 wide range of lifetimes, placing stringent
 limits on the \h\ production process.
 This result, in conjunction with the result
from experiment BNL E888~\cite{belz}, completely rules out the model
proposed in  
 reference~\cite{donoghue} for all $\Delta S = 1$ transitions.

We thank D.~Ashery,  F.S.~Rotondo and
A.~Schwartz for their insightful comments.
We gratefully acknowledge the support and effort of the Fermilab
staff and the technical staffs of the participating institutions for
their vital contributions.  This work was supported in part by the U.S. 
Department of Energy, The National Science Foundation and The Ministry of
Education and Science of Japan. 
%In addition, A.R.B., E.B. and S.V.S. 
%acknowledge support from the NYI program of the NSF; A.R.B. and E.B. from 
%the Alfred P. Sloan Foundation; E.B. from the OJI program of the DOE; 
%K.H., T.N. and M.S. from the Japan Society for the Promotion of
%Science.  

%
%	Figures
%

\end{document}